\documentclass[conference]{IEEEtran}
\IEEEoverridecommandlockouts
\usepackage{cite}
\usepackage{amsmath,amssymb,amsfonts}
\usepackage{algorithm}  
\usepackage{algpseudocode}
\usepackage{graphicx}
\usepackage{textcomp}
\usepackage{xcolor, amsmath}
\usepackage{amsmath}
\usepackage{bbm}
\usepackage{bbold}

\DeclareMathOperator*{\argmax}{argmax}

\def\BibTeX{{\rm B\kern-.05em{\sc i\kern-.025em b}\kern-.08em
    T\kern-.1667em\lower.7ex\hbox{E}\kern-.125emX}}

\title{Fast Initial Access with Deep Learning for Beam Prediction in 5G mmWave Networks \\
\thanks{
This effort is supported by the U.S. Army Research Office under contract W911NF-20-P0035. The content of the information does not necessarily reflect the position or the policy of the U.S. Government, and no official endorsement should be inferred.
}
}

\author{\IEEEauthorblockN{ Tarun S. Cousik, Vijay K. Shah, and Jeffrey H. Reed}
\IEEEauthorblockA{\textit{Wireless@VT, Dept. of ECE, Virginia Tech} \\
Blacksburg, USA \\
\{tarunsc, vijays, reedjh\}@vt.edu}
\and
\IEEEauthorblockN{Tugba Erpek and Yalin E. Sagduyu}
\IEEEauthorblockA{\textit{Intelligent Automation, Inc.}\\
Rockville, USA \\
\{terpek, ysagduyu\}@i-a-i.com}}

\begin{document}

\maketitle

\begin{abstract}
This paper presents DeepIA, a deep learning solution for faster and more accurate initial access (IA) in 5G millimeter wave (mmWave) networks when compared to conventional IA.
By utilizing a subset of beams in the IA process, DeepIA removes the need for an exhaustive beam search thereby reducing the beam sweep time in IA. 
A deep neural network (DNN) is trained to learn the complex mapping from the received signal strengths (RSSs) collected with a reduced number of beams to the optimal spatial beam of the receiver (among a larger set of beams).  In test time, DeepIA measures RSSs only from a small number of beams and runs the DNN to predict the best beam for IA. We show that DeepIA reduces the IA time by sweeping fewer beams and significantly outperforms the conventional IA's beam prediction accuracy in both line of sight (LoS) and non-line of sight (NLoS) mmWave channel conditions. 
\end{abstract}


\section{Introduction}\label{sec:Intro}
The shift to operate on millimeter wave (mmWave) frequencies in order to satisfy increasing bandwidth requirements has forced the adoption of highly directional antennas and/or arrays (HDAs) to combat the path loss associated with these high frequencies. Initial access (IA) in mmWave systems serves the purpose of orienting the HDAs of two or more radio devices (which are unaware of their relative positions) to point at one another to establish the initial connection \cite{1,2,3,4}. IA is a key component in 5G systems to establish the initial connection of a mobile user and the cellular network. 

The adoption of HDAs introduces an alignment window challenge in IA. Beam alignment becomes more difficult when narrower beams are used and needs to be repeated more frequently due to potential blockage effects in mmWave systems that may otherwise result in loss of beam alignment. However, IA is an expensive process in terms of transmit power and computational costs involved, and if not properly orchestrated, may not leave enough time for communications. 
Phase 1 of the 5G wireless standard allows a maximum of 64 synchronization signal blocks (SSBs) in the spatial plane.  The time allocated for this sweep of 64 beams is $5$ ms and this process is repeated every $5$-$20$ ms~\cite{tripathi}. However, the coherence time (over which the channel remains constant) experienced by mobile platforms is much smaller than $20$ ms. Therefore, novel methods are needed to decrease the time it takes for the IA thereby improving the connection time to start communications. 

The IA time consists of two components, (i) time for beam sweeping (measuring the received signal strengths (RSSs) for different beams) and (ii) time for beam prediction (identifying the beam for a given transmitter-receiver pair to communicate with). Since the beam sweep time dominates the overall IA time, it is essential to improve the IA time by utilizing fewer beams. However, a conventional beam sweeping (CBS) approach that selects the best beam based on RSSs from a reduced set of beams cannot be highly accurate. This is because CBS cannot predict any of the beams that are not used in beam sweeping and the subset of beams that are selected for beam sweeping need not contain the best beam. However, the mapping from the RSSs measured for a subset of beams to the best beam is a complex process due various channel, antenna, and network topology effects. This limitation motivates us to understand if and how a data-driven approach could be leveraged in predicting beams that are not part of the beam sweeping process. 

Machine learning provides automated means to learn from spectrum data and perform complex tasks such as spectrum sensing \cite{spectrumsensing}, signal classification \cite{signalclass}, anti-jamming \cite{TCCN}, and waveform design \cite{MIMOWav}. Supported by recent advanced in algorithmic techniques and computational resources, deep learning has emerged as a viable solution to capture high-dimensional representations of spectrum data \cite{DL}. In this paper, we propose DeepIA as a deep learning solution to reduce the beam sweep time by measuring RSSs from only a subset of all 
available beams and mapping them to the best selection from the entire set of beams. 
The deep neural network (DNN) trained in DeepIA learns to associate the correct beam (sector) between the transmitter and receiver with the RSS perceived by the receiver. Hyperparameters of the DNN are carefully selected to avoid both underfitting and overfiting. For both LoS and NLoS mmWave channel conditions, we show that compared to CBS, DeepIA not only provides a faster IA scheme but also a more reliable one in terms of accuracy. For example, DeepIA predicts the optimal beam with close to $100\%$ accuracy in LoS conditions by sweeping only $6$ out of $24$ beams. The accuracy of conventional beam sweeping for the same setting is limited to $24\%$.   

In addition, DeepIA reduces the overall computational and operational costs by reducing the number of transmissions/beams each transmitter has to sweep through during IA. Consequentially, it also decreases the temporal and spatial footprint of the signal in the open environment which reduces the interference from/to other ongoing transmissions (such as in spectrum sharing scenarios) as well as the probability of detection/intercept. Thus, DeepIA could potentially improve the resiliency against out-network interference and jamming.

  The rest of the paper is organized as follows. 
Related work is described in Section \ref{sec:relatedwork}. The system model is described in Section \ref{sec:systemmodel}. Algorithmic solution is presented in Section \ref{sec:proposed_solution}. Performance analysis with both LoS and NLoS mmWave channel models are provided in Section \ref{sec:perfanalysis} that also expands on the expected computational times for beam prediction and beam sweeping when using embedded platforms. Finally, the conclusion is presented in Section \ref{sec:conclusion}.


\section{Related Work} \label{sec:relatedwork}
Conventional IA involves exhaustive or iterative beam searches. In both cases, a predefined number of beams divide the azimuth plane into sectors, with each beam covering a unique sector\cite{ia1, ia2, ia3}. One possible implementation of the exhaustive search involves the transmitter cycling through every sector while the receiver runs in a quasi-omnidirectional mode and listens for the transmitted signal. The receiver (e.g., a user equipment (UE) in 5G) records the RSS  for each transmit beam and finally sends back the best performing beam to the transmitter (e.g., gNodeB in 5G). The transmitter then turns on this particular beam, and the receiver performs the aforementioned beam sweeping process to determine its best sector. This exhaustive search is computationally inefficient. 
Unlike omnidirectional sub-6GHz systems, this IA process has to be periodically repeated in mmWave systems in order to ensure that there is no misalignment between transmitter-receiver pairs over time. 

Another version of the exhaustive search runs narrow transmitter beams against narrow receiver beams\cite{ia1}. Iterative search uses a combination of wide and narrow beams on the transmitter and receiver to minimize the computational cost of the exhaustive search at the expense of decreasing the detection accuracy  \cite{ia2}. In addition, since only a partial set of the antenna elements are used to create the beams, it can decrease the range over which these wide beams can be used.  A hybrid IA process is introduced in \cite{ia2}, where first the iterative process is performed and then the receiver sends the uplink signals in the best beam and transmitter finds its best narrow beam after cycling through all its narrow beams against the best receiver beam. This hybrid IA process performs equally with the iterative search in terms of accuracy but utilizes approximately $86$-$90\%$ of the time resources. 
\cite{ia3} has proposed to use multi-beam analog beamformers which simultaneously overlay several narrow beams in the spatial domain for IA. While the transmit time is reduced, computational and temporal requirements for the back-end processing needed to distinguish the beams and detection accuracy remain unclear. \cite{ia4} has reduced the average discovery time by leveraging knowledge obtained from real time arrival statistics of incoming users. In this paper, we use only a reduced set of narrow beams without any prior information, and train a DNN to predict the best beam from a larger set. This approach sustains high accuracy in beam prediction while reducing the time for beam sweeping.    

\section{System Model} \label{sec:systemmodel}

\begin{figure}[h!]
	\centerline{\includegraphics[width=0.95\linewidth]{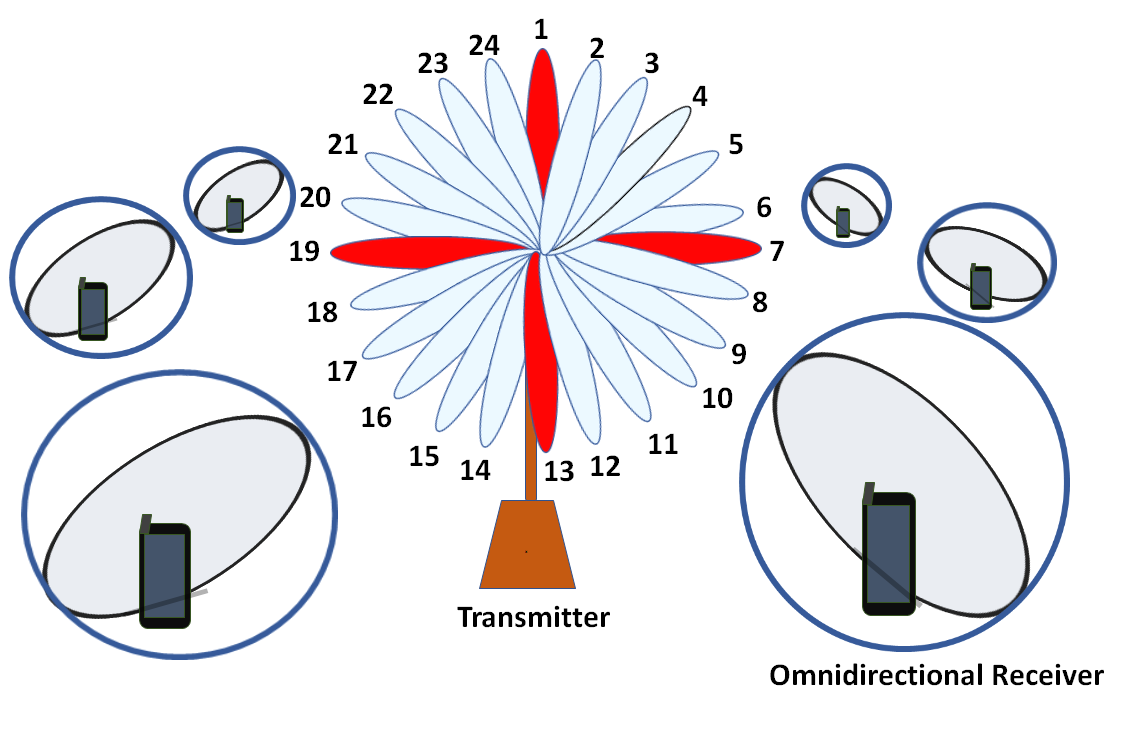}}
	\caption{System model.}
	\label{fig:sys}
\end{figure}


We consider a 5G mmWave network that consists of a directional transmitter and omnidirectional receivers as shown in Figure \ref{fig:sys}. Without loss of generality, we consider a 2D plane where the transmitter is located at $(0, 0)$ and $R$ receivers are uniformly randomly distributed in a square area. No receiver is placed within a $1$ meter radius of the transmitter to support the close intercept (CI) model used in Section \ref{subsec:mmWaveCha}. 

\subsection{Transmitter Antenna Array Characteristics}
\begin{figure}[h!]
\begin{center}
\includegraphics[scale=0.56]{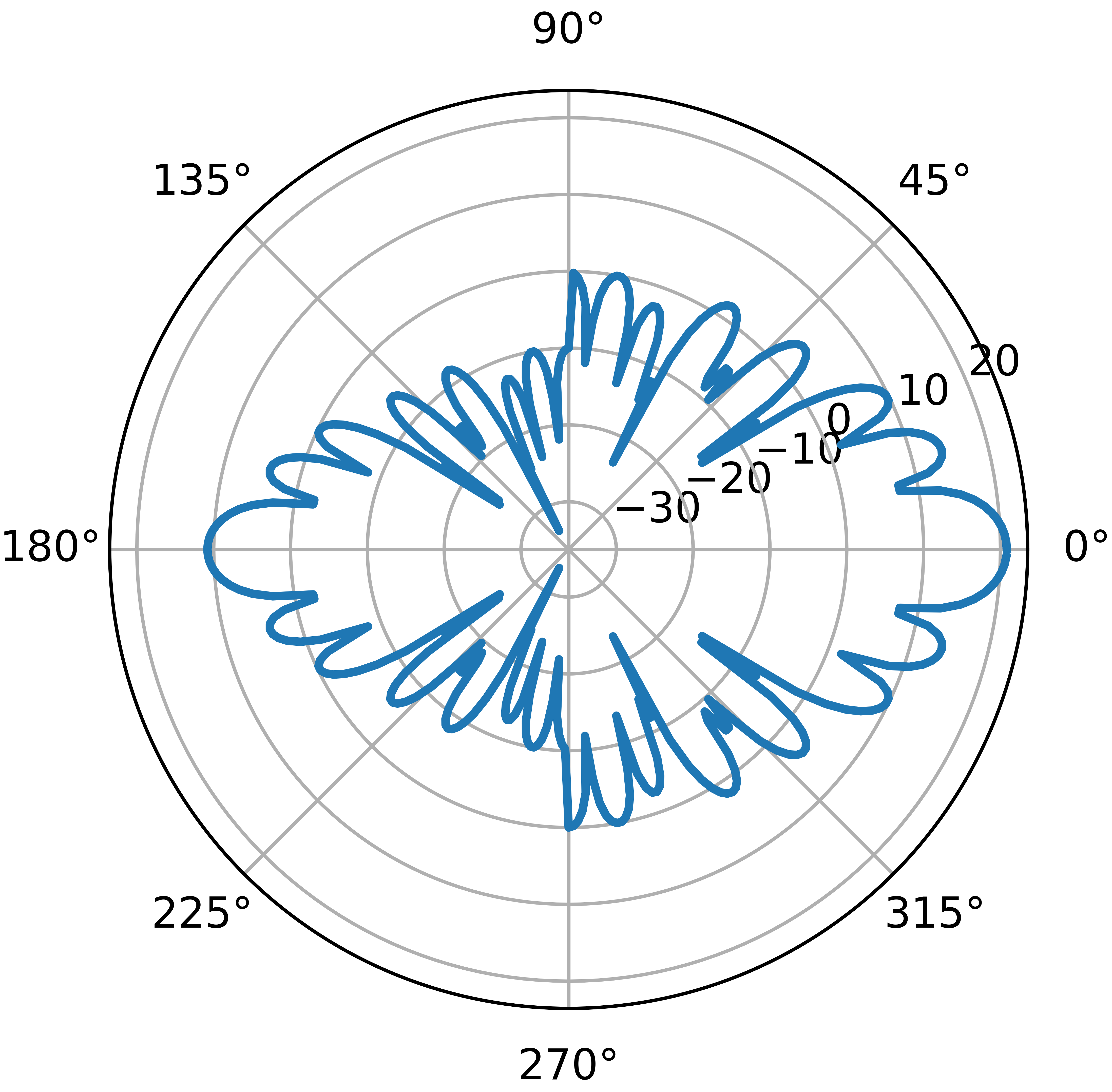}
\caption{Antenna pattern (gain values in dB).}
\label{fig:AP}
\end{center}
\end{figure}

We design a $10 \times 10$ antenna array at the transmitter which generates a beam width of about $15 ^{\circ}$ using the standard planar array formulation \cite{balanis}. This array points at an azimuth angle $\phi = 0^{\circ}$ and an elevation angle $\theta = 90^{\circ}$. Since we consider a 2-D scenario, we take the azimuthal slice of the array factor at $\theta = 90^{\circ}$. We then subtract $10$ dB from this standard array slice for azimuth angles between $180^{\circ}$ and $360^{\circ}$. This modification serves two purposes. First, antenna designers routinely modify the array/antenna patterns to ensure minimal backlobe radiation. The goal is to avoid transmitting in unnecessary directions, reduce electromagnetic interference to back-end electronics and to avoid jamming users in the backlobe's direction. 
Second, using a standard $10 \times 10$ planar array pattern severely affected the performance of CBS because of pattern symmetricity. When the array factor of the backlobe has the same magnitude as that of the front lobe this causes confusion on which beam to pick. The modified array pattern used in this paper is shown in Fig. \ref{fig:AP}. This  modified array pattern is consistently used in all of the sectors instead of simulating the array pattern for every sector individually to reduce the computational overhead during training. 

\subsection{Channel Models} \label{subsec:mmWaveCha} 
In our simulations, we consider LoS and NLoS mmWave channels. 
We use the CI path loss model, where at distance $d$, the path loss in dB is given by $PL(d) = PL(d_{0}) + 10n \log_{10}\frac{d}{d_{0}} + X_{\sigma} $, where $PL(d_{0}) = 10\log_{10}\left(\frac{4 \pi d_{0}}{\lambda}\right)^2$ and $d_0 = 1$ m. $n$ is the best fit minimum mean square error path loss exponent (PLE), $n = 4.5$ for NLoS conditions and $n = 1.9$ for LoS conditions. $X_{\sigma}$ is the shadow factor representing large scale signal fluctuations 
and is a zero-mean Gaussian random variable with a standard deviation of $10$ for NLoS conditions and $1.1$ for LoS conditions \cite{mmWave_rap}.



\subsection{Overview of the Baseline IA \label{subsec:baseline}}
The baseline that we refer to as CBS uses exhaustive search \cite{ia1}, 
with a directional transmitter and an omnidirectional receiver. CBS exceeds both the iterative and hybrid approaches in terms of detection accuracy\cite{ia2}. We define $\mathcal{N} = \{1,...,N\}$ as the set of all $N$ beams that the transmitter can sweep. 
Let $r_k$ denote the $k$th receiver, where $k\in\{1,2,3...R\}$, and $RSS_{ik}$ denote the RSS corresponding to the $i$th beam between the receiver $r_k$ and the transmitter.
In CBS (detailed in Algorithm \ref{P0_algo}), all $N$ beams in $\mathcal{N}$ are swept serially and the $\hat{i}_k$th beam which provides the highest RSS is selected, i.e., for receiver $r_k$, the selected beam is $\hat{i}_k = \argmax\limits_{i \in \mathcal{N}} RSS_{ik}$.

\begin{algorithm} [h!]
\label{alg_1}
	\textbf{Input:} Receiver $r_k$\\
	\textbf{Output:} Selected beam $\hat{i}_k$
	\begin{algorithmic}[1]
	\For{$i \in \mathcal{N}$} 
	\State Measure $RSS_{ik}$
	\EndFor
	\State $\hat{i}_k = \argmax\limits_{i \in \mathcal{N}} RSS_{ik}$
	\end{algorithmic}  
	\caption{Baseline CBS Algorithm}
	\label{P0_algo}  
\end{algorithm}	


\subsection{Overview of DeepIA}\label{subsec:overview_DeepIA}

DeepIA uses the global information of receiver positions during the training phase. Given that the receiver's position is known to the transmitter, the best beam is calculated based on the angle between them as explained later in Section \ref{subsec: beam_mapping}. The DNN is trained by feeding the RSS values from a set $\mathcal{M}$ of $M$ beams as the input, where $\mathcal{M} \subseteq \mathcal{N}$.
 
As the DNN, we use a feed-forward neural network (FNN) architecture. The input layer consists of $M$ neurons, each receiving input from one beam from $\mathcal{M}$. To illustrate this idea, the subset $\mathcal{M} = \{1,7,13,19\}$ (i.e., $M = 4$) is shown in red in Fig. \ref{fig:sys}. At the output layer, there are $N$ output neurons corresponding to the $N$ beams and their magnitudes denote the respective probabilities of being the optimal beam. During training, the labels, i.e., the best beam for a given transmitter and receiver position, are determined based on the angle between the transmitter and receiver instead of the measured RSS values. 
The subsets of $M$ beams are uniformly sampled from $N$ and used during both the training and test times in our simulations. 


\section{Deep Learning Framework of DeepIA} \label{sec:proposed_solution}
This section describes the details of DeepIA's deep learning framework; namely, DNN architecture, beam mapping and label generation, and training and testing phases. 

\subsection{Deep Neural Network Architecture} \label{subsec:NNarchitecture}

The DNN architecture of DeepIA consists of seven layers including the input and output layers. The input layer has $M$ neurons, and the output layer has $N$ neurons with each neuron providing a likelihood score. The hidden layers have $64$, $128$, $256$, $128$ and $64$ neurons, respectively. These hidden layer structures are determined after tuning the hyperparameters based on the training and validation performance. In simulations, we use $N =24$ and $M$ is either $2$, $4$, $6$, $8$, $12$, or $24$.
The hidden layers are activated by ReLu, Tanh, Sigmoid, ReLU and Tanh functions, respectively. A Softmax activation function is used at the output layer. The number of neurons in each layer and the activation functions used are shown in Table \ref{table:DeepIANNArch}. 
Each layer's output except that of the output layer is batch normalized before passing it on to the next layer. The backpropagation algorithm is used to train the DNN by minimizing the cross entropy loss function with a learning rate of $10^{-3}$. The Adam optimizer is used in converging to the minima of the loss function \cite{adam}.


\begin{table}[h!]
\caption{Deep Neural Network Architecture.}
\centering
{\small
\begin{tabular}{l|l|l}
\textbf{Layers} & \textbf{\# neurons} & \textbf{Activation function}\\ \hline  \hline
Input & $M = 2, 4, \dots,$ or $24$ &  $-$  \\ \hline 
Dense 1 & 64 & ReLu \\ \hline
Dense 2 & 128 & Tanh \\ \hline
Dense 3 & 256 & Sigmoid \\ \hline
Dense 4 & 128 & ReLu \\ \hline
Dense 5 & 64 & Tanh \\ \hline
Output  & $N = 24$ & Softmax \\
\end{tabular}
}
\label{table:DeepIANNArch}
\end{table}


\subsection{Beam Mapping and Label Generation} \label{subsec: beam_mapping}


 The process for beam mapping and label generation, which is a key requirement for generating the training dataset, is described below.

\begin{figure}[h!]
\begin{center}
\includegraphics[scale=0.45]{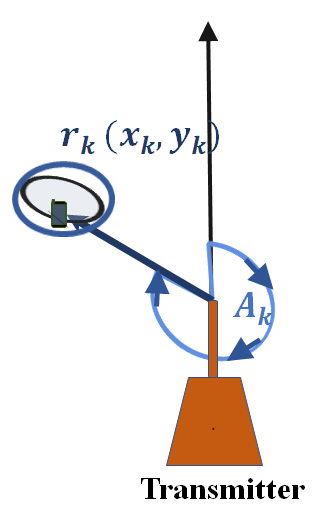}
\caption{Angle between the transmitter and the receiver.}
\label{fig:angle}
\end{center}
\end{figure}

As shown in Fig. \ref{fig:angle}, the angle between the $k$th receiver ${r_k}$ and the transmitter is denoted as $A_k$. This angle represents the relative orientation between any receiver position and transmitter. $A_k$ can take values between $0^{\circ}$ and $359.99 ^{\circ}$, and is rounded to the lowest integer. Transmitter $A_k$ for every receiver $r_k$ is then mapped to its corresponding beam as follows. When $N= 24$, angles $((-8^{\circ}, 7^{\circ}],(7^{\circ},22^{\circ}],(22^{\circ},37^{\circ}],\dots,(337^{\circ},352^{\circ}])$ are mapped to Beams $1, 2, ..., 24$ respectively. Note that $-a^{\circ}$ is the same as $360^{\circ}-a^{\circ}$. This is performed by setting $i^*_k$ =  $\lceil\frac{A_k +8}{15}\rceil$, where $i^*_k$ represents the true beam sector receiver $r_k$ should correspond to. $i^*_k$ also corresponds to the labels that we use against our training, test and validation features. In addition, we use $i^*_k$ in evaluating the accuracy of both the CBS and DeepIA. 

\subsection{Deep Neural Network Training} \label{subsec:training}


DeepIA uses RSS values from $M$ beams as inputs. The features are converted to linear units and normalized with respect to the maximum value. This ensures that the DNN has input values that range from $0$ to $1$. Note, the values in Tables \ref{tab:trainingdata} and \ref{tab:testingdata} are written in dB scale to help visualize the procedure, in reality they are normalized to values between 0 and 1.
The DNN is trained to learn to map $\left\{RSS_{ik}\right\}_{i \in \mathcal{M}}$ values to the corresponding $i^*_k$ values. 
We show a visualization of what the DNN training set would look like in Table \ref{tab:trainingdata}. The receiver column corresponds to the index of different receivers. We separate the $R$ receivers and their corresponding data into $3$ sets for training, validation and testing. The number of receivers used for training, validation and testing are given by $R_{tr}$, $R_{val}$ and $R_{te}$, respectively. Therefore,  $R = R_{tr} + R_{val} + R_{te}$. The output of training for beam set $\mathcal{M}$ is the trained DNN model $\texttt{DeepIA}_{\mathcal{M}}: \left\{ RSS_{ik} \right\}_{i \in \mathcal{M}} \rightarrow \hat{i}_k \in \mathcal{N}$. 
\begin{table}[h!]
    \centering
    \caption{An example of training data entries.}
    \begin{tabular}{c|c|c|c|c}
         Receiver & Beam $1$ & \dots & Beam $M$ & True Beam \\ \hline \hline
         1 & -30 & \dots & -46 & 1 \\ \hline
         2 & -51 & \dots & -24 & 5 \\ \hline
         - & - & \dots & - & - \\ \hline
         $R_{tr}$ & -65 & \dots & -68 & 2 \\ 
    \end{tabular}
    \label{tab:trainingdata}
\end{table}

Six DNN models are trained for $M = 2,4,6,8,12,$ and $24$ and number of epochs used to train these six models are $35,55,65,70,75,$ and $90$, respectively. We observe that DeepIA learns faster as we increase the number of beams. So, the optimal number of epochs to train decreases with an increase in $M$. A learning rate of $10^{-3}$ is used. A batch size of $1024$ is consistently maintained for all the models. The hyperparameters of the DNN are determined after evaluating the validation accuracy and epochs where the training loss saturates.




\subsection{Deep Neural Network Testing} \label{subsec:testing}
The RSS values from the same $M$ beams used during training are measured and fed to the DNN model for receiver indices that were never seen before. The index of the output with the highest probability is selected as the best spatially oriented beam. Table~\ref{tab:testingdata} shows an example test input and output with $\mathcal{M}$ inputs. The predicted beam should match exactly with the true beam for a correct prediction. In cases where they do not match up such as that of receiver $r_{k}$, DeepIA makes an error in its prediction. 

DeepIA (detailed in Algorithm \ref{P0_algo2}) takes as input the subset of beams $\mathcal{M} \subseteq \mathcal{N}$ for any receiver $r_k \in R$, and measures $RSS_{ik}$ for each beam $i \in \mathcal{M}$ (see lines 1-3 in Algorithm \ref{P0_algo2}). Then, DeepIA utilizes the trained DNN architecture  $\texttt{DeepIA}_{\mathcal{M}}$ to predict the best beam $\hat{i}_k$ (see line 4 in Algorithm \ref{P0_algo2}).

\begin{algorithm} [h!]
\label{alg_1}
	\textbf{Input:} Receiver $r_k$, $\texttt{DeepIA}_{\mathcal{M}}$\\
	\textbf{Output:} Predicted beam $\hat{i}_k$
	\begin{algorithmic}[1]
	\For{$i \in \mathcal{M}$}
	\State Measure $RSS_{ik}$
	\EndFor
	\State $\hat{i}_k = \texttt{DeepIA}_{\mathcal{M}}\left(\{RSS_{ik}\}_{i \in \mathcal{M} }\right)$
	\end{algorithmic}  
	\caption{DeepIA Algorithm}
	\label{P0_algo2}  
\end{algorithm}	

\begin{table}[h!]
    \centering
    \caption{An example of test data entries}
    \begin{tabular}{c|c|c|c|c|c}
         Receiver & Beam $1$ & \dots & Beam $M$ & Predicted Beam& True Beam \\ \hline \hline
         1 & -35 & \dots & -46 & 1 & 1 \\ \hline
         2 & -51 & \dots & -64 & 5 & 5 \\ \hline
         - & - & \dots & - & - \\ \hline
         $r_{k}$& -27 & \dots & -71 & 4 & 5\\ \hline
         - & - & \dots & - & - \\ \hline
         $R_{te}$ & -25 & \dots & -28 & 2 & 2 \\ 
    \end{tabular}
    \label{tab:testingdata}
\end{table}


\section{Performance Analysis} \label{sec:perfanalysis}

\subsection{Simulation Setting}

 We consider a two-dimensional network scenario where transmitters and receivers all lie on the same plane. The location of the transmitter is fixed at $(0,0)$ and the receiver's X and Y positions each take random values derived from sampling a uniform random variable  between $-25$ m and $25$ m. This bounds the simulation cell to an area of $50 \times 50$ m. The transmit power is set at $20$ dB.  
 
 A total of $10^6$ receiver positions are generated, which act as data samples. Out of total $10^6$ data samples, $65\%$ are used to train DeepIA, $15\%$ of data samples are used for validation, and $20\%$ of data samples are used for testing. 
 



\subsection{Comparison Approaches}
We evaluate DeepIA against CBS in terms of both accuracy and beam prediction time, for varying subset of beams (out of total beams).

\begin{itemize}
    \item \textbf{Prediction Accuracy:}  
    The prediction accuracy is computed as the ratio of receivers for which the predicted beam matches the true beam, namely
    \begin{equation*}
        \text{Prediction Accuracy} [\%] = \frac{\sum_{k=1}^{R} \mathbb{1}( i^*_k = \hat{i}_{k})}{R} \times 100,
    \end{equation*}
    where $\mathbb{1}$ is the indicator function ($\mathbb{1}(E) =1$ if $E$ holds and $0$, otherwise).
    \item \textbf{IA Time:}  The IA time is the total time incurred in predicting the correct beam for a certain transmitter-receiver position. It constitutes two parts described below. 
    \begin{enumerate}
        \item \textit{Beam Sweep Time:} This refers to the time taken to sweep a desired number of beams. The beam sweep time is directly proportional to the number of beams swept. For instance, Phase 1 of 5G NR\cite{tripathi}, allows a duration of $5$ ms for sweeping across a total of $64$ beams. Going by these numbers, it should take $1.875$ ms to sweep $24$ beams, $0.9375$ for $12$ beams,  $0.675$ ms for $8$ beams, and so on. In simulations, we consider the same subset of beams to sweep for both DeepIA and CBS. Therefore, we compare the accuracy performance under the same beam sweep time.   
        
        \item \textit{Beam Prediction Time:} This refers to the time it takes to process the collected RSS data and predict the beam it belongs to. In CBS, this involves calculating a maximum operation on the RSS collected from $N$ beams. In DeepIA, this involves calculating the time it takes to run the DNN model using $M$ inputs in order to predict the beam a given receiver belongs to. 
    \end{enumerate}
\end{itemize}

\subsection{Simulation Results}
In this section, we compare the performance of DeepIA against the baseline CBS, in terms of prediction accuracy and IA time.



\subsubsection{Prediction Accuracy Analysis} 
\begin{figure}[ht]
    \centering
\includegraphics[scale = 0.6]{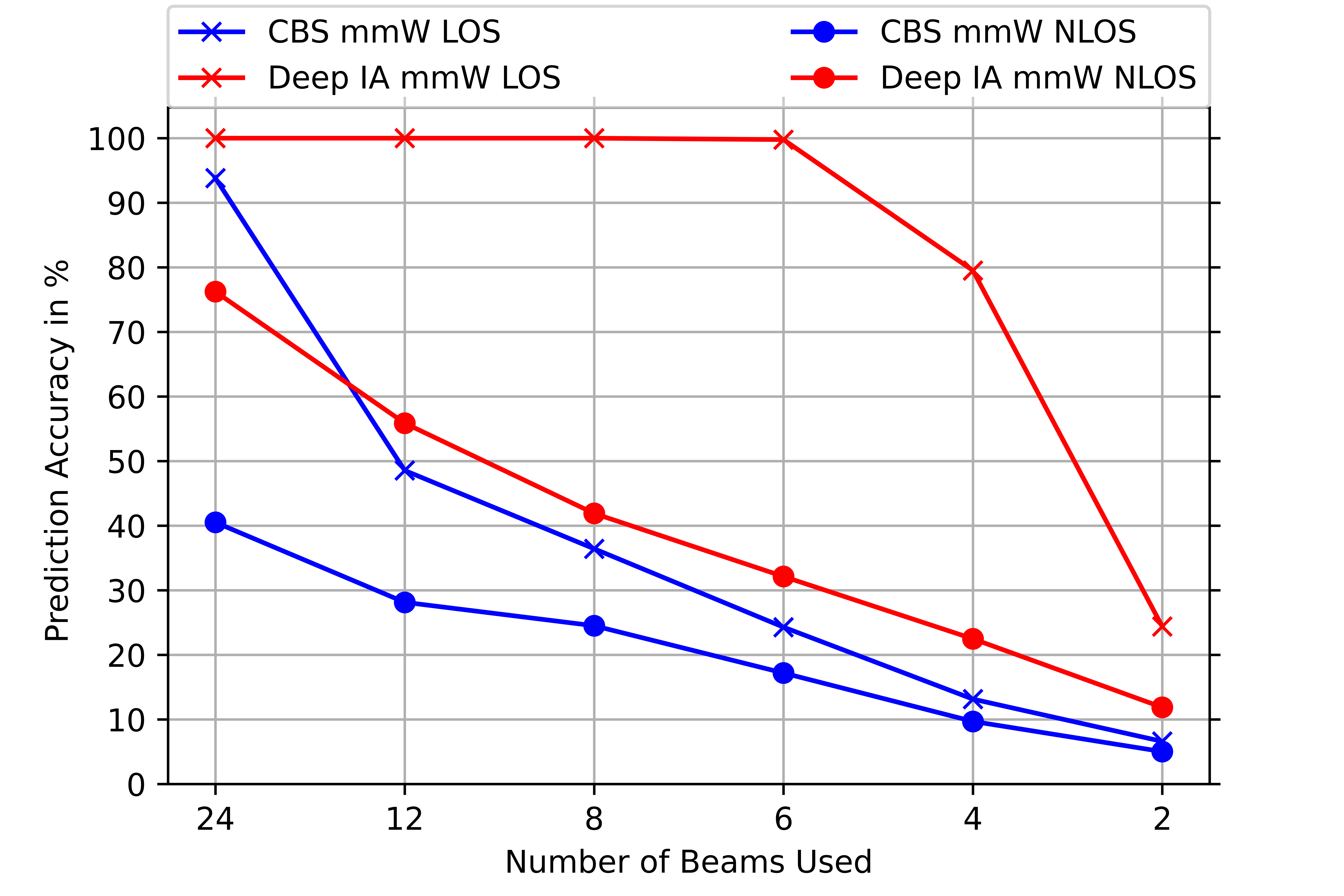}
\caption{DeepIA vs. CBS for LoS and NLoS mmWave path loss models.}
\label{fig:NN_result1}
\end{figure}

The prediction accuracy is shown in Fig. \ref{fig:NN_result1}. Since the exhaustive beam search in CBS relies on scanning the entire spatial plane, its accuracy continuously deteriorates as fewer beams are utilized. On the other hand, DeepIA demonstrates a more robust and reliable prediction accuracy over the range of beams that are swept. For example, in the LoS condition even setting $M =6$ achieves close to $100\%$ prediction accuracy. Therefore, adding more beams after a certain point adds very little benefit in terms of accuracy in the DNN approach under LoS conditions. 

In the LoS mmWave channel, we observe a deterioration in performance of DeepIA when 6 or fewer beams are utilized. There is a minimum number of beams that are required in order to predict highly accurate results and this minimum is reached at 8. 
With severe shadow fading in the NLoS mmWave channels DeepIA's performance deteriorates. 
Nonetheless, DeepIA still outperforms CBS, especially until $8$ beams are used. In a time-sensitive scenario, it would be practical to use DeepIA with $8$ beams, since it gives $100\%$ accuracy in LoS conditions and performs equally compared to an exhaustive $24$ beam CBS while utilizing approximately $33\%$ of the total compute time. If performance is critical and if the channel is predominantly NLOS, it makes sense to use DeepIA with $24$ beams. 
Thus, DeepIA provides a network operator a prediction accuracy vs. beam alignment time trade off to work with.

\subsubsection{Beam Prediction Time Analysis}
For practical implementation aspects, 5G allows $5$ ms periodicity by default to complete the beam sweeping procedure in the IA. However, coherence time over which the channel remains unchanged is much smaller for highly mobile users communicating at high frequencies. Coherence time $T_C$ is inversely proportional to the maximum Doppler Spread $D_S = \frac{f_c v}{c}$, where $f_c$ is the center frequency, $v$ is the speed of mobile user, and $c$ is the speed of light \cite{Tse}. As an example, consider $f_c = 28$ GHz. If the user moves with walking speed (e.g., $v = 1.4$ m/s), then $T_C$ is less than $7.7$ ms. On the other hand, if the user moves with a car speed of $v = 25$ m/s, then $T_C$ is less than $0.43$ ms. Therefore, it is important to make the beam prediction decision within this time frame.

To better understand the computation time in practice, we consider running the DNN configuration of DeepIA on an embedded platform. For that purpose, the procedure of \cite{Soltani2019} to convert the trained software model for a deep neural network to the Field Programmable Gate Array (FPGA) code is followed and computational aspects of running DeepIA's DNN are evaluated for the FPGA implementation. Vivado Design Suite \cite{Vivado} is used to simulate and then synthesize the FPGA code with the 16-bit implementation for Xilinx UltraScale FPGA. 


We evaluate the processing time associated with beam sweeping and beam prediction as follows. Beam sweeping involves a pilot tone transmitted from each beam at the transmitter one at a time to the receiver. The receiver measures the RSS value for each beam. This process is repeated $N$ times for CBS algorithm and $M$ times for DeepIA. The selection of the best beam from RSS values can either be performed at the transmitter or receiver. In the former case, the receiver transmits the measured RSS value back to the transmitter after every beam transmission as a feedback. In the latter case, the receiver predicts the best beam itself and transmits a single feedback. Assuming it takes $t_s$ to receive a tone signal at the receiver, the beam sweeping time is reduced from $Nt_s$. to $Mt_s$. The duration of beam sweeping is linearly proportional to the number of beams swept and is much larger than the beam prediction time that we discuss next. 

The duration of beam prediction however, depends on whether CBS or DeepIA is used. Suppose $M = 6$ beams are swept to fix the beam sweeping time for both CBS and DeepIA. The duration to take the maximum operation in CBS is $MT$ with one comparator, $ \left(\left\lceil \frac{M}{2}\right\rceil +1\right)T$ with two comparators, and so on. Hence, CBS selects its beam in at most $0.06$ $\mu$s. On the other hand, DeepIA runs the RSSs from $6$ beams through its  DNN. The latency to process one input sample (namely, $6$ RSSs) through the DNN architecture given in Table \ref{table:DeepIANNArch} is $6.9$ $\mu$s. The processing power spent by the FPGA is $1.03$ W. Since the beam prediction time in both CBS and DeepIA is at most at microsecond level, the time for beam sweeping dominates the total time spent for IA. Therefore, DeepIA reduces the time for IA significantly and leaves more time for data communications.



\section{Conclusion} \label{sec:conclusion}
In this paper, we developed the deep learning-based fast IA solution, DeepIA, for 5G mmWave networks. DeepIA collects measurements only from a small number of beams to predict the best beam for data communications. While outperforming the CBS that searches for the best signal strength in all beams, DeepIA can successfully capture the complex patterns of transmitter-receiver locations and beam patterns, and can predict the best beam with very high accuracy even when a small number of beams are swept in the IA. Compared to CBS, DeepIA significantly reduces the time for IA and leaves more time for communications. Our simulation results show that DeepIA is able to predict the optimal beam with nearly $100\%$ accuracy even when $6$ out of $24$ beams are used in LoS conditions, while the beam prediction accuracy of CBS drops to around $24\%$ with the same number of beams.

\end{document}